# Ferroelectric metals in van der Waals bilayers

Jiagang Zhang,[1] Ting Zhang[1*]

School of Physics and Technology, University of Jinan, Jinan 250022, People's Republic of China

The combination of metallicity and ferroelectricity breaks traditional boundaries, paving new avenues for innovative electronic materials and devices. This breakthrough is particularly notable, as metallicity and ferroelectricity have traditionally been considered mutually exclusive physical properties. In this work, starting with non-polar metallic single layers, we propose a general scheme for designing 2D ferroelectric metals (FEMs) based on van der Waals interaction. By first-principles calculations, we further substantiate the feasibility of the design scheme in real materials such as FeSe and H-MnTe$_2$. Notably, this scheme unveils unique metallic ferroelectricity, characterized by reversing polarization through interlayer sliding. Furthermore, the combination of inherent magnetism with sliding ferroelectricity leads to multiferroicity. The investigated design scheme and observed phenomena have broad applicability across 2D materials. Our results not only pave the way for research in 2D FEMs but also offer promising prospects for foundational studies of coupled physical phenomena in 2D lattices.

Ferroelectric materials exhibit switchable bistable states, each characterized by an opposite spontaneous electric polarization reversible through an external electric field [1-4]. Recently, two-dimensional (2D) ferroelectric materials have garnered interest because of their novel physical properties and potential applications in non-volatile memory devices [2,5]. According to Neumann's principle, ferroelectrics typically belong to one of the 10 polar point groups, characterized by a unique rotation axis, and lacking both the inversion center and the mirror plane [6]. However, many 2D materials with triangular or hexagonal lattices cannot support spontaneous vertical polarization due to their inversion or mirror symmetries. Recently, the introduction of sliding ferroelectricity has injected new vitality into the field of 2D ferroelectric materials, significantly broadening the scope of the 2D ferroelectric family [7-10]. The van der Waals stacking typically breaks the crystal symmetry, thus ferroelectric polarization can also occur in certain non-polar point groups. This phenomenon has been theoretically and experimentally confirmed in various materials, including multi-layer h-BN and H-$MoS_2$ [11-13].

In traditional perceptions, ferroelectricity and metallicity are considered mutually exclusive physical characteristics, as the conductive electrons in metals can effectively screen internal electric fields, thereby inhibiting the manifestation of ferroelectricity [14]. However, in 1965, Anderson and Blount posited the potential existence of ferroelectric metallic materials (FEMs) [14,15]. Some materials have been proposed as FEMs, such as $LiOsO_3$ [15,16], but its polarization cannot be switched by an external electric field. It is noteworthy that in the field of 2D materials, the coexistence of ferroelectricity and metallicity seems possible. In 2D materials, the confinement of electrons within the plane allows an external electric field to penetrate and reverse the vertical polarization [17], which opens a new direction for the study of 2D FEMs. While theoretical predictions have been made regarding the ferroelectric metallic behavior in 2D materials such as CrN [18] and $MIMIIP_2X_6$ [19], to date, only multilayer 1T'-$WTe_2$ has been successfully confirmed as a FEM in experiments [3,20,21]. Consequently, the research of 2D FEMs remains in its early stages, with a relative scarcity of suitable FEMs.

In this work, we propose a universal design scheme for 2D FEMs, starting with a metallic single-layer lattice and constructing a bilayer lattice through van der Waals stacking. This scheme exploits lattice rotation and translation to break inversion and mirror symmetries, thereby introducing sliding ferroelectricity into metallic bilayer lattice. Employing first-principles calculations, we verify the applicability of this design scheme in materials such as bilayer FeSe and H-$MnTe_2$. Moreover, the integration of inherent magnetism with sliding ferroelectricity induces multiferroicity. Our investigation into this design scheme and the phenomena observed demonstrate wide applicability across 2D materials. These findings not only broaden the scope of 2D FEMs but also pave new paths for further exploration in this field.

All calculations based on the Density Functional Theory (DFT) are performed using the Vienna Ab-initio Simulation Package (VASP) [22]. The exchange-correlation effects are treated using the Generalized Gradient Approximation (GGA) parameterized by Perdew, Burke, and Ernzerhof (PBE) [23]. The ion-electron interactions are handled with the Projector Augmented-Wave (PAW) method [24,25]. The Monkhorst-Pack *k*-point grid is set to 19 × 19 × 1. The kinetic energy cutoff is set to 450 eV. Full relaxation is applied to all structures until the force on each atom is less than 0.01 eV/Å. The electronic convergence criterion is set to $1 \times 10^{-5}$ eV. To avoid spurious interactions between periodic images, a vacuum space of at least 15 Å is introduced. The van der Waals interactions are treated using the DFT-D3 method in all calculations. The Non-Equilibrium Bardeen (NEB) method is employed to calculate the ferroelectric transition path and energy barrier [26]. Dipole moment correction is applied to evaluate the polarizations [27].

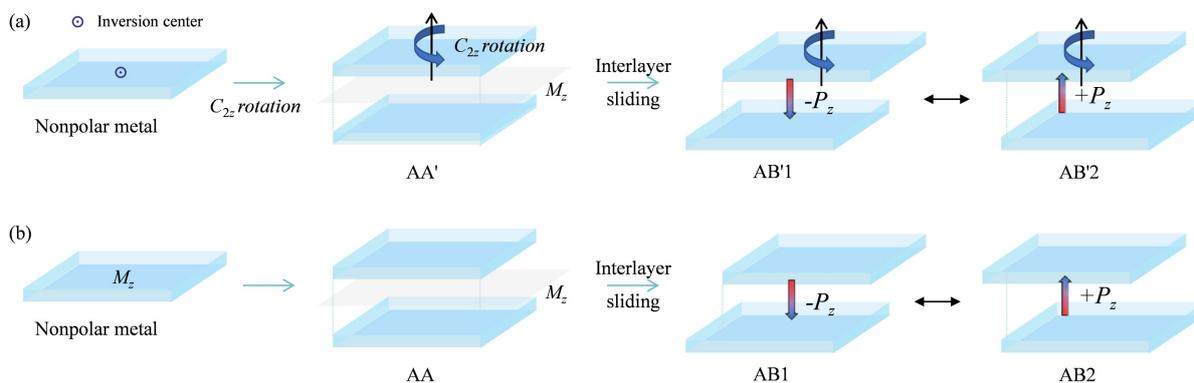

**Figure 1** Schematic diagrams of design schemes for realizing 2D FEMs based on van der Waals interactions.

In physics, achieving ferroelectricity in metals is a challenging task due to the inherent contradiction between ferroelectricity and metallicity [14]. To address this, we propose a general mechanism based on the theory of sliding ferroelectricity—achieving metallic ferroelectricity in 2D van der Waals stacked metals through interlayer sliding. The van der Waals stacked metals exhibit vertical polarization since electrons are confined within the slab and nonconductive along the vertical direction [20]. By allowing for interlayer sliding, two metallic ferroelectric phases with opposite polarization can switch with low energy barrier. Our proposed mechanism begins with non-polar metallic single layer. We explore two distinct stacking patterns: (a) the AA' pattern and (b) the AA pattern; see **Figure 1**.

(a) In the case of non-polar metallic single layers associated with space groups $P\bar{3}$, $P\bar{3}1m$ and $P\bar{3}m1$, to break their inversion centers, we adopt the AA' stacking pattern; see **Figure 1(a)**. This configuration is achieved by rotating one metallic single layer by 180° before stacking it onto the other layer. Although the resulting AA' pattern lacks an inversion center, it still retains mirror symmetry $M_z$, which obstructs the emergence of vertical electric polarization. To overcome this, we introduce interlayer sliding $t_{1//}\left[-\frac{2}{3},-\frac{1}{3},0\right]$, $t_{2//}\left[\frac{1}{3},-\frac{1}{3},0\right]$ or $t_{3//}\left[\frac{1}{3},\frac{2}{3},0\right]$ ($-t_{1//}$, $-t_{2//}$ or $-t_{3//}$), transforming the AA' pattern into the AB'1 (AB'2) pattern. The elimination of both inversion and mirror symmetries in these configurations results in opposite vertical polarizations. The in-plane confinement of electrons allows the vertical polarization to exist and can be switched by a vertical electric field. These two patterns represent two distinct ferroelectric states that can be switched via interlayer sliding.

(b) For non-polar metallic single layers belonging to space group $P\bar{6}m2$, characterized by mirror symmetry $M_z$, we employ the AA stacking pattern, as shown in **Figure 1(b)**. Here, two single layers are directly stacked to form a bilayer lattice. By inducing interlayer sliding $t_{1//}$, $t_{2//}$ or $t_{3//}$ ($-t_{1//}$, $-t_{2//}$ or $-t_{3//}$), its $M_z$ symmetry is broken, leading to the AB1 (AB2) ferroelectric state. The breaking of vertical symmetry in the two patterns triggers a charge redistribution, which generates vertical electric polarization. This process introduces ferroelectricity into the van der Waals stacked metallic lattices, thereby expanding the potential applications of FEMs in 2D materials.

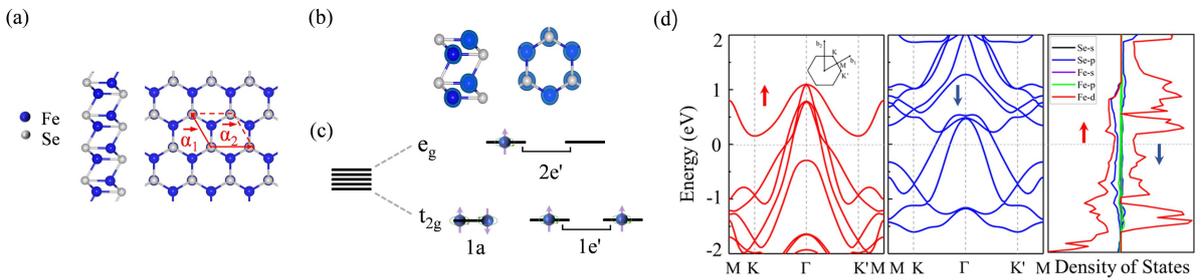

**Figure 2** (a) Crystal structures of single-layer FeSe. (b) Spin charge density distributions of single-layer FeSe. (c) The splitting of $d$ orbitals in single-layer FeSe. (d) Band structures and projected density of states of single-layer FeSe.

2D FeSe emerges as a candidate material to illustrate the design schemes in case (a).

Single-layer FeSe consists of two buckled FeSe sublayers, which are interconnected by Fe-Se bonds; **see Figure 2(a)**. In this arrangement, the Fe and Se atoms of the upper layer are positioned directly above the Se and Fe atoms of the lower layer, respectively. Each unit cell comprises two Fe and two Se atoms, with a lattice constant of 3.67 Å, and is classified under the space group P$\bar{3}$m1. The in-plane and out-of-plane Fe-Se bond lengths measure 2.40 Å and 2.45 Å, respectively. A notable feature of single-layer FeSe is its absence of vertical electric polarization, a characteristic inherently tied to its centrosymmetric structural configuration that leads to the neutralization of polarization. To delve deeper into the stability of single-layer FeSe, we conducted both phonon dispersion analysis and ab initio molecular dynamics (AIMD) simulations. The phonon spectrum displays an absence of significant imaginary frequencies, indicating the dynamic stability of single-layer FeSe [15]; see **Figure S1(a)**. AIMD simulations at 300 K show that single-layer FeSe maintains its ground-state configuration without discernible structural deformations, indicating its thermal stability [28]; see **Figure S1(b)**.

After investigating the structure and stability of single-layer FeSe, we further explored its magnetic properties. We observed that the ground state of single-layer FeSe exhibits ferromagnetic coupling. The results show a magnetic moment of 3 $u_B$ on Fe atoms, contrasting with the negligible moments on the connected Se atoms. The spin charge density distributions confirm that the magnetic moment is primarily localized on Fe atoms, as shown in **Figure 2(b)**. The Fe-$d$ orbitals are split into $e_g$ (2e') and $t_{2g}$ (1a and 1e') orbitals in single-layer FeSe; see **Figure 2(c)**. The 1a orbital is fully occupied, and the 1e' orbital houses two electrons in separate orbitals, while the 2e' orbital contains one electron, resulting in the observed total magnetic moment of 3$u_B$ in single-layer FeSe. Additionally, analysis of the band structures and projected density of states reveals the absence of a band gap, indicating the metallicity of single-layer FeSe.

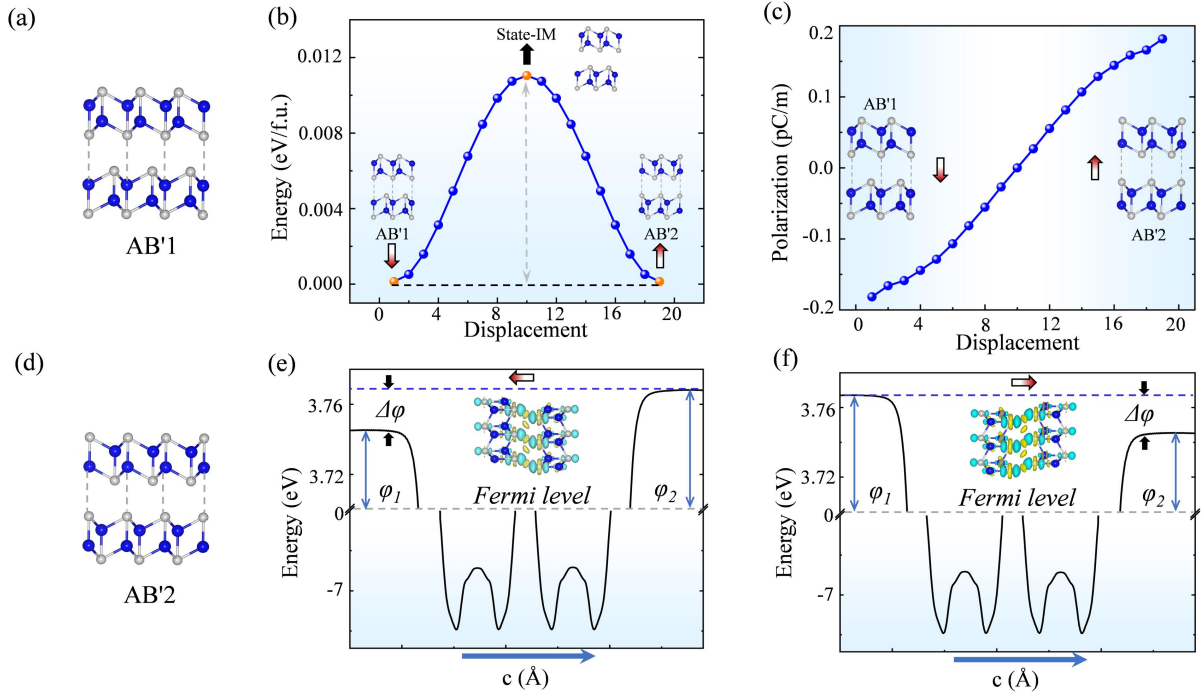

**Figure 3** Crystal structures of the (a) AB'1 and (d) AB'2 patterns of bilayer FeSe. (b) Energy profiles associated with the ferroelectric transition in bilayer FeSe. The inset in (b) shows the intermediate state without polarization (State-IM). (c) Changes in the vertical electric polarization of bilayer FeSe during the ferroelectric transition. Plane averaged electrostatic potentials of the (e) AB'1 and (f) AB'2 patterns of bilayer FeSe along the *z* direction. Insets in (e) and (f) display the differential charge densities of the AB'1 and AB'2 patterns, respectively.

According to the proposed mechanism, we constructed a bilayer FeSe with the AB'1 stacking pattern, as shown in **Figure 3(a)**. The Se atoms of the upper layer are positioned directly above the Fe atoms of the lower layer. This configuration belongs to the P3m1 space group, lacking inversion center and $M_z$ symmetry. The lattice constant for bilayer FeSe is 3.65 Å, with an interlayer spacing of 3.01 Å. Both interlayer and intralayer couplings in bilayer FeSe exhibit FM interactions. The energy of interlayer FM coupling is 0.009 eV/f.u., lower than that of interlayer AFM coupling. **Figure S2** displays the band structures and density of states for bilayer FeSe, confirming its alignment with our design principle as a 2D FEM.

In the AB'1 pattern of bilayer FeSe, the Fe atoms of the lower layer align directly beneath the Se atoms of the upper layer. This distinctive non-equivalent bilayer configuration results in an interlayer imbalance, which subsequently causes electron redistribution between the two layers, thereby inducing vertical electric polarization [3]. The vertical polarization in the AB'1 pattern is 0.18 pC/m. This polarization is further substantiated by differential charge density and plane averaged electrostatic potential analyses [29]. As shown in **Figure 3(e)**, the positive discontinuity ΔV = 0.02 eV between the upper and lower vacuum layers distinctly indicates the spontaneously directed downward vertical polarization in the AB'1 pattern. Moreover, the differential charge density displays the asymmetry. Notably, the vertical electric polarization in bilayer FeSe is not screened due to its in-plane metallic nature and the vertical confinement of electrons [20].

By introducing reverse interlayer sliding, we obtained the AB'2 pattern, as shown in **Figure 3(d)**. In the AB'2 pattern, the interface Se atoms from the lower layer are positioned beneath the Fe atoms from the upper layer, reversing the atomic alignment observed in the AB'1 pattern of bilayer FeSe. Consequently, the AB'2 pattern generates a vertical electric polarization pointing upwards. The value of electric polarization is 0.18 pC/m, but the direction is opposite to that of the AB'1 pattern. We investigated the plane averaged electrostatic potentials and differential charge density of the AB'2 pattern to further corroborate this point. **Figure 3(f)** clearly illustrates the spontaneous vertical electric polarization in the AB'2 pattern, as evidenced by a negative discontinuity ΔV =− 0.02 eV between the vacuum levels of the upper and lower layers, with the polarization direction pointing upward. Additionally, the differential charge density exhibits significant asymmetry.

The existence of electric polarization in metallic bilayer FeSe is not sufficient to guarantee ferroelectricity unless this polarization can be switched. To this end, we investigated the feasibility of polarization switching in bilayer FeSe. **Figure 3(b)** shows the energy pathway for ferroelectric switching between the AB'1 and AB'2 patterns using the nudge-elastic-band (NEB) method [30]. Because of the rotational symmetry $C_{3z}$, the AB'1 pattern can transform into the AB'2 pattern through interlayer sliding along $\left[-\frac{2}{3},\frac{2}{3},0\right]$, $\left[\frac{4}{3},\frac{2}{3},0\right]$ or $\left[-\frac{2}{3},-\frac{4}{3},0\right]$. The intermediate state (state-IM), characterized by the space group Abm2 [see inset of **Figure 3(b)**], lacks polarization due to the presence of a glide plane along the $c$-axis. The energy barrier for ferroelectric switching is determined to be 0.011 eV/f.u.. **Figure 3(c)** illustrates the corresponding ferroelectric polarization as a function of the step number. Notably, this energy barrier is higher than that of bilayer T'-WTe$_2$ [20,21] and h-BN [11-13], while being lower than that of In$_2$Se$_3$ (0.066 eV/f.u.) [17,31]. This barrier supports the feasibility of ferroelectric switching between the AB'1 and AB'2 patterns in bilayer FeSe. Therefore, based on the proposed mechanism, the AB'1 and AB'2 patterns of bilayer FeSe can be regarded as two switchable metallic ferroelectric states.

Because of the presence of vertical electric polarization, the magnetic moments on the Fe atoms in the upper and lower layers are not entirely equivalent. Interlayer sliding-induced changes in the ferroelectric state also affect the distribution of magnetic moments on Fe atoms [32]. When switching from the AB'1 to the AB'2 pattern, the reversal of polarization leads to a redistribution of magnetic moments, making those in the lower layer slightly greater than those in the upper layer, in contrast to the AB'1 pattern. This phenomenon enables the control of magnetic moment distribution in bilayer FeSe through ferroelectricity, resulting in the magnetoelectric coupling. This indicates that bilayer FeSe is also a multiferroic metal.

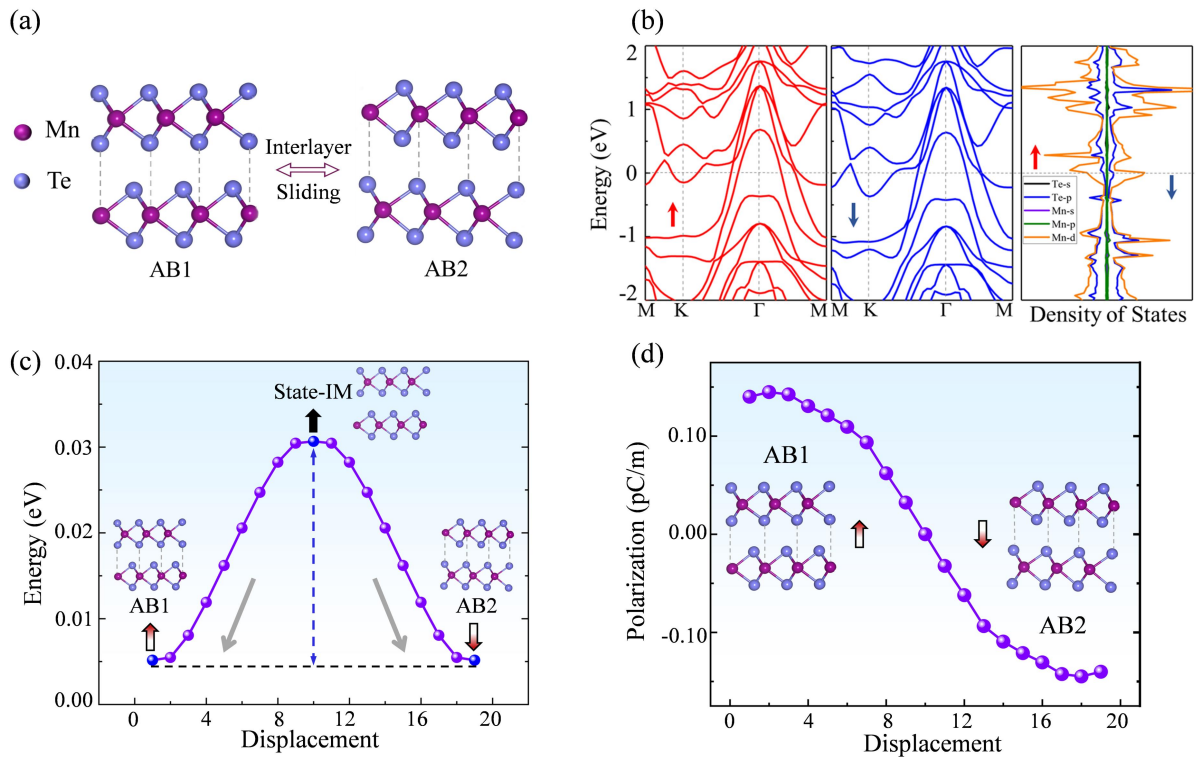

**Figure 4** (a) Crystal structures of the AB1 and AB2 patterns of bilayer H-MnTe$_2$. (b) Band structures and projected density of states of bilayer H-MnTe$_2$. (c) Energy profiles associated with the ferroelectric transition in bilayer H-MnTe$_2$. The inset in (c) shows the non-polar state-IM. (d) Changes in the vertical electric polarization of bilayer H-MnTe$_2$ during the ferroelectric transition.

Next, our focus turns to case (b), with a detailed examination into its realization in the real material H-MnTe$_2$. The single-layer H-MnTe$_2$ belongs to the $P\bar{6}m2$ space group, where Mn atoms form a hexagonal lattice, each Mn atom coordinating with six Te atoms to create an octahedral structure. The phonon spectrum and AIMD simulations of the single-layer H-MnTe$_2$ are shown in **Figure S3(a)** and **(b)**. The absence of imaginary phonon frequencies in the single-layer H-MnTe$_2$ indicates its high dynamic stability [15]. Moreover, its structure remains undistorted at 300 K, indicating high thermal stability [28].

The magnetic ground state of single-layer H-MnTe$_2$ exhibits FM coupling, with the magnetic moment primarily localized on the Mn atoms. Each Mn atom possesses the electronic configuration of $3d^54s^2$. Upon donating three valence electrons to the surrounding Te atoms, the oxidation state of Mn ions transitions to +3, leading to the electronic configuration of $3d^34s^0$. In an octahedral crystal field, the Mn $d$ orbitals split into two sets: the higher energy doublet $e_g$ orbitals and the lower energy triplet $t_{2g}$ orbitals; see **Figure S3(c)**. Thus, the three valence electrons occupy the $t_{2g}$ orbitals, while the $e_g$ orbitals remain unoccupied. This arrangement results in a magnetic moment of 3 $\mu_B$ per Mn atom in single-layer H-MnTe$_2$.

In accordance with the proposed scheme, we construct the AB1 and AB2 patterns of bilayer H-MnTe$_2$, as shown in **Figure 4(a)**. The lattice constants for both patterns are 3.67 Å. Our results indicate that the interlayer exchange interactions in bilayer H-MnTe$_2$ support AFM coupling. The energy of interlayer AFM coupling is 0.005 eV/f.u. lower than that of interlayer FM coupling. The band structure and density of states (DOS) are presented in **Figure 4(b)**, indicating that bilayer MnTe$_2$ exhibits metallicity. The AB1 and AB2 patterns belong to the space group of P3m1, lacking both an inversion center and $M_z$ symmetry. In the AB1 pattern of bilayer H-MnTe$_2$, the Te atom of the upper layer is directly above the Mn atom of the lower layer, leading to vertical electric polarization, and vice versa for the AB2 pattern. This conclusion is further substantiated by analyses of differential charge density and plane averaged electrostatic potential. A negative (positive) discontinuity $\Delta V = -0.018$ (0.018) eV between the vacuum levels of the upper and lower layers serves as a clear indicator of the spontaneous vertical electric polarization, with an upward (downward) orientation, within the AB1 (AB2) pattern; see **Figure S4**. Furthermore, the differential charge density exhibits the asymmetry. The value of electric polarization in the AB1 and AB2 patterns of bilayer H-MnTe$_2$ is calculated to be 0.14 pC/m.

In accordance with the proposed scheme, the AB1 and AB2 patterns of bilayer H-MnTe$_2$ can be regarded as two sliding ferroelectric states. The pathway for ferroelectric switching between these patterns is shown in **Figure 4(c)**. Because of the rotational symmetry $C_{3z}$, the AB1 pattern can transition into the AB2 pattern through interlayer sliding along $\left[-\frac{2}{3}, \frac{2}{3}, 0\right]$, $\left[\frac{4}{3}, \frac{2}{3}, 0\right]$ or $\left[-\frac{2}{3}, -\frac{4}{3}, 0\right]$. The state-IM, characterized by space group Abm2, exhibits non-polarity due to a glide plane along the $c$-axis. **Figure 4(d)** illustrates the ferroelectric polarization as a function of the step number. The ferroelectric energy barrier for bilayer H-MnTe$_2$ is determined to be 0.026 eV/f.u., lower than that of In$_2$Se$_3$ (0.066 eV/f.u.) [17,31]. This barrier facilitates the possibility of ferroelectric transition between the AB1 and AB2 patterns in bilayer H-MnTe$_2$. Additionally, interlayer sliding leads to a modification in the magnetic moment distribution on Mn atoms, indicating the presence of magneto-electric coupling [8]. In summary, our investigation reveals the magnetic, ferroelectric, and metallic properties of bilayer H-MnTe$_2$, confirming its interlayer sliding-induced switchable metallic ferroelectricity.

In conclusion, we introduce a novel design scheme for 2D FEMs through van der Waals interactions. Our first-principles calculations validated this approach in real materials like bilayer FeSe and H-MnTe$_2$, unveiling a unique metallic ferroelectricity with switchable vertical polarization via interlayer sliding. This method not only demonstrates a low energy barrier but also integrates inherent magnetism, leading to the emergence of multiferroicity. Applicable to a broad spectrum of 2D materials, our findings herald a new opportunity in the exploration of 2D FEMs.

## AUTHOR INFORMATION

Corresponding Authors

*E-mail: sps_zhangt@ujn.edu.cn (T.Z.).

ORCID

Ting Zhang: 0009-0008-5670-9955

**Notes**

The authors declare no competing financial interest.

## SUPPORTING INFORMATION

Supporting Information Available: details regarding the phonon spectrum and the molecular dynamics simulation of single-layer FeSe and H-MnTe$_2$, the band structures and the projected density of states of bilayer FeSe, and the differential charge densities of bilayer H-MnTe$_2$.